\documentclass[preprint2]{aastex}

\shorttitle{Radiation Mechanism of the Soft Gamma-ray Pulsar PSR B1509}
\shortauthors{Wang, Takata, \& Cheng}

\begin{document}

\title{Mechanism of the X-ray and Soft Gamma-ray Emissions from High Magnetic Field Pulsar : PSR B1509-58}
\author{Y. Wang, J. Takata, and K.S. Cheng} 
\affil{Department of Physics, University of Hong Kong, Pokfulam Road, Hong Kong}
\email{yuwang@hku.hk (YW); takata@hku.hk (JT); hrspksc@hkucc.hku.hk (KSC) }

\begin{abstract}
We use the outer gap model to explain the spectrum and the energy dependent light curves of the X-ray and soft $\gamma$-ray radiations of the spin-down powered pulsar PSR B1509-58. In the outer gap model, most pairs inside the gap are created around the null charge surface and the gap's electric field separates the two charges to move in opposite directions. Consequently, the region from the null charge surface to the light cylinder is dominated by the outflow current while that from the null charge surface to the star is dominated by the inflow current. We suggest that the viewing angle of PSR B1509-58 only receives the inflow radiation. The incoming curvature photons are converted to pairs by the strong magnetic field of the star. 
The X-rays and soft $\gamma$-rays of PSR B1509-58 result from the synchrotron radiation of these pairs.
Magnetic pair creation requires a large pitch angle, which makes the pulse profile of the synchrotron radiation distinct from that of the curvature radiation. We carefully trace the pulse profiles of the synchrotron radiation with different pitch angles. We find that the differences between the light curves of different energy bands are due to the different pitch angles of the secondary pairs, and that the second peak appearing at $E>10$MeV comes from the region near the star, where the stronger magnetic field allows pair creation to happen with a smaller pitch angle.

\end{abstract}


\section{Introduction}
PSR B1509-58 (hereafter PSR B1509) is a unique subject of the multi-wavelength study of the high energy radiation mechanism of the spin-down powered pulsar. It was discovered in soft X-ray band by Einstein (Seward \& Harnden 1982) and soon was detected in the radio band (Manchester et al. 1982) thirty years ago. PSR B1509 is a young (about 1600 years old) spin-down powered pulsar with a period of about 150 ms and a surface magnetic field of $\sim 1.5\times{}10^{13}$G, which is much stronger than those of most of the canonical pulsars ($\sim{}10^{12}$G). After being discovered, its radiation was observed by many X-ray and soft $\gamma$-ray telescopes: EXOSAT (Trussoni et al. 1990), Ginga (Kawai et al. 1992), BASTE and OSSE on the CGRO (Ulmer et al. 1993; Matz et al. 1994), SIGMA (Laurent et al. 1994), Welcome-1 (Gunji et al. 1994), ROSAT (Greiveldinger et al. 1995), RXTE (Marsden et al. 1997; Rots et al. 1998; Livingstone \& Kaspi 2011), ASCA (Saito et al. 1997), COMPTEL (Kuiper et al. 1999) and BeppoSAX (Cusumano et al. 2001). These observations show that, from 0.1 keV to 10 MeV, PSR B1509 has a wide single peak, whose shape can be described by two Gaussian functions, and COMPTEL shows that, when $E>10$MeV, another peak shows up near the peak of radio emission. After 2008, the bottleneck in $\gamma$-ray astrophysics was broken by $Fermi$ and AGILE. In 2010, these two $\gamma$-ray telescopes provided the observed data of PSR B1509 above 30MeV (Abdo et al. 2010; Pilia et al. 2010). The soft $\gamma$-ray radiation and the second peak found by COMPTEL have been confirmed by them.

PSR B1509 is unique among the known $\gamma$-ray spin-down powered pulsars for its very soft $\gamma$-ray spectrum and a very strong surface magnetic field. It is suggested that there may be a new class of spin-down powered pulsar with high magnetic field ($>$10$^{13}$G) and soft $\gamma$-ray radiation, called soft $\gamma$-ray pulsar (Pilia \& Pellizzoni 2011). Also, there is at least another pulsar that is similar to PSR B1509, PSR J1846-0258, a young pulsar with $P \sim 324$ms and surface magnetic field $B_{s} \sim 4.9\times{}10^{13}$G, which cannot be detected by $Fermi$ but is seen by X-ray telescopes (Kuiper \& Hermsen 2009; Parent et al. 2011). Therefore, a multi-wavelength study of the radiation mechanism of PSR B1509 can be a bridgehead to investigate this possible new class of pulsars.

Recent observations show that two kinds of pulsars, the spin-down powered pulsar with $B_s\sim{}10^{12}$G and the magnetar with $B_s\sim{}10^{14}-10^{15}$G, have no clear boundary between them. It is believed that the radiation of a magnetar (AXP or SGR) is powered by the energy stored in the super strong magnetic field with $B_s>10^{14}$G (Duncan \& Thompson 1992), however, two low magnetic field soft $\gamma$-ray repeaters, SGR 0418+5729 with $B_s<7.5\times{}10^{12}$G (Rea et al. 2010) and J1822.3-1606 with $B_s\simeq2.7\times{}10^{13}$G (Rea et al. 2012), have been found, which challenge the requirement of a super strong magnetic field of the existing theoretical models. Furthermore, radio emission, which is usually regarded as a characteristic behavior of a spin-down powered pulsar, has been found in three magnetars, XTE J1810-197 (Camilo et al. 2006), 1E 1547-5408 (Camilo et al. 2007) and PSR J1622-4950 (Levin et al. 2010). On the other hand, PSR J1846-0258, a spin-down powered pulsar with $B_s\sim{}4.9\times{}10^{13}$G, is found to have a magnetar-like outburst (Kuiper \& Hermsen 2009). So, the study of the radiation mechanism of the very high magnetic field spin-down powered pulsars, and the connection between the canonical spin-down powered pulsar and the magnetar, can help us to understand the physics of the two groups of neutron star. As a family member of the high magnetic field spin-down powered pulsars, PSR B1509 may provide some insights for understanding the differences between typical pulsars and magnetars.

As a known $\gamma$-ray pulsar before the $Fermi$ era, PSR B1509 has been interpreted by two main radiation scenarios, the polar cap model (Ruderman \& Sutherland 1975; Daugherty \& Harding 1982) and the outer gap model (Cheng, Ho \& Ruderman 1986; Romani 1996; Hirotani 2006). For the polar cap model, where the $\gamma$-ray photons are emitted near the stellar surface, the soft $\gamma$-ray photons are generated via the photon splitting process $\gamma\to\gamma\gamma$ caused by a strong magnetic field (Harding \& Baring 1997). In the point of view of the outer gap model, where the $\gamma$-ray radiation comes from the outer magnetosphere, Zhang and Cheng (2000) assumed PSR B1509  to be another Crab pulsar and explained the X-ray to soft $\gamma$-ray radiation of this pulsar as the synchrotron radiation of the secondary pairs created by photon-photon pair creation. Since the light cylinder of PSR B1509 is much larger than that of the Crab pulsar, the inverse Compton component of PSR B1509 is much weaker than that of the Crab. They predicted that the GeV energy flux produced by inverse Compton scattering is nearly three orders of magnitude weaker than that of the MeV energy flux. However this predicted energy flux is nearly two orders of magnitude lower than the recent observations by AGILE (Pilia et al. 2010) and Fermi (Abdo et al. 2010). It is interesting to note that the observed total radiation power of PSR B1509 is less than 1\% of its spin-down power, which is much lower than the efficiency predicted by most theoretical models (e.g. Zhang \& Cheng 1997; Takata, Wang \& Cheng 2010). Furthermore, the light cylinder of PSR B1509 is so large as to make it very difficult to attenuate all of the GeV curvature photons emitted from the outer gap.

Cheng, Ruderman \& Zhang (2000) showed that most electron/positron pairs created inside the outer gap are produced near the null charge surface, the strong electric field inside the gap separates the two charges to move in opposite directions. Therefore from the null charge surface to the light cylinder the radiation is mainly outward, and from the null charge surface to the star the radiation is mainly inward. Since the light cylinder of PSR B1509 is much bigger than that of the Crab pulsar, most outflow curvature photons can  avoid photon-photon pair creation. If the line of sight is in the direction of the outgoing radiation beam, the spectrum of PSR B1509 should be a characteristic pulsar spectrum, namely a power law with exponential cut-off as predicted by Wang, Takata \& Cheng (2010). Obviously this conflicts with the observed properties of PSR B1509. In this paper we propose that the viewing angle of PSR B1509 is in the direction of the incoming radiation beam, where the curvature photons with hundreds MeV are converted into pairs by the strong magnetic field. The observed X-rays and soft $\gamma$-rays are the synchrotron radiation of these pairs. It is also important to note that the electric field from the null charge surface decreases rapidly towards the inner boundary (e.g. Hirotani et al. 2003; Takata, Chang \& Cheng 2007; Tang et al. 2008), therefore the incoming radiation flux is at least an order of magnitude weaker than that of the outgoing flux.

Most of these incoming curvature photons emitted by canonical pulsars with a surface magnetic field $\sim 10^{12}$G will be converted into pairs and hence they become invisible. 
{ On the other hand if they are not absorbed they may appear near the phase position of the radio pulse. Figure~\ref{lc_colmap} gives the pulse phases and viewing angles of the incoming curvature photons, which are represented by the gray lines. 
The pulse phase of the center of the polar cap is set to be 0, because the radio emission is believed to come from the polar cap. If the viewing angle is larger than the inclination angle of the pulsar and there is no magnetic pair creation,
the gray lines will make a peak near the phase 0. }
There are some clear evidences for the existence of this incoming curvature photon beam in the energy range of a few hundred MeV in gamma-ray millisecond pulsars whose surface magnetic field is 
of order of $B_s\sim{}10^{8}$G (Abdo et al. 2010b) but this mangetic field strength is not strong enough to convert photons of several hundred MeV into pairs. 
Even very strong multiple fields may exist on the surface of millisecond pulsars but they decrease much more rapidly than the dipolar field. 
The incoming photons of several hundred MeV can escape from the magnetic pair creation process unless they come closer than a distance of 2-3 stellar radii (Takata, Wang \& Cheng 2010).

If the pairs are generated by photon-photon pair creation, which is used to explain the radiation of the Crab pulsar in Tang et al. (2008), the pulse phases of the peaks of X-ray radiation are the same as those of the $\gamma$-ray radiation, because of the small pitch angles of the secondary pairs.  On the other hand, the magnetic pair creation requires the photon to penetrate with a large angle to the magnetic field, therefore the incoming $\gamma$-ray photons have to go inwards to obtain enough $B_{\perp}$, and the secondary pairs obtain larger pitch angles than those from photon-photon pair creation. These pitch angles make the pulse profile of the synchrotron radiation of these secondary pairs different from that of the curvature radiation of the primary particles. It has been found that the non-thermal X-ray pulse profile of the spin-down powered pulsar can be different from that of the $\gamma$-ray, such as PSR J0659+1414 and PSR J1420-6048 (Weltevrede et al. 2010). This indicates that the radiation mechanism discussed in this paper can be further applied to explain the non-thermal X-ray radiation of other $\gamma$-ray pulsars.

In section 2, we describe our theoretical model. Section 3 gives the details of the calculation of the pulse profile of the synchrotron radiation emitted by the secondary pairs generated by the magnetic pair creation process. In section 4, we present our calculation of the spectrum and the energy dependent light curves of PSR B1509. We discuss some special features in the observed spectrum and light curves in section 5. A brief summary is given in section 6.

\section{Theoretical Model}
\subsection{The structure of the gap}

\begin{figure}[t]
\plotone{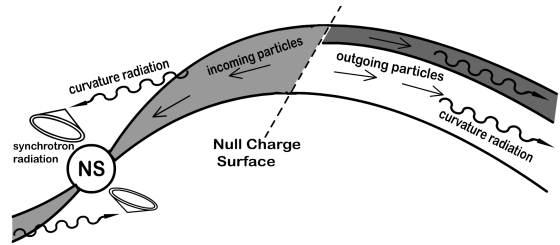}
\caption{The sketch of the structure of the gap. Outside the null charge surface the structure of the gap can be simplified as a two-layer structure. From the null charge surface to the stellar surface, the accelerating electric field is screened out.}
\label{gap_structure}
\end{figure}

Figure~\ref{gap_structure} illustrates the structure of our theoretical model. From the stellar surface to the light cylinder, along the magnetic field line, the emission region has different structures on the two sides of the null charge surface. For PSR B1509, only the radiation from the region inside the null charge surface is received by the viewing angle while that from the region outside the null charge surface is missed. Outside the null charge surface, the particles go outwards along the open field lines, and the structure of the gap can be simplified as a two-layer structure (Wang et al. 2010, 2011) which contains a main acceleration region and a screening region. The main acceleration region, where the charge density is much lower than the Goldreich-Julian charge density, $\rho_{GJ}\sim-\frac{\vec{B}\cdot\vec{\Omega}}{2\pi{}c}$ (Goldreich \& Julian 1969), is the source of the curvature photons with $E>1$GeV, and a strong accelerating electric field is formed from such deviation. Above the main acceleration region, the screening region has a charge density much larger than $\rho_{GJ}$, which screens out the electric field in the trans-field direction. In this region, the charged particles near the edge between the two layers can be accelerated to emit curvature photons of hundreds of MeV. Inside the null charge surface, the electric field is reduced and there is no two-layer structure. 

In the direction along the magnetic field line, the accelerating electric field is screened out in a similar way as in the trans-field direction outside the null charge surface. The potential that generates the accelerating electric field in the gap is given by the Poisson equation $\nabla^2\Phi'=-4\pi{}(\rho-\rho_{GJ})$. In the region of two-layer structure, according to this Poisson equation, in the trans-field direction, a strong accelerating electric field is formed in the region where the charge density is close to zero (the main acceleration region), and screened out in the region where the charge density is larger than $\rho_{GJ}$ (the screening region). The average charge density of the two-layer region is $\sim{}0.5\rho_{GJ}$ (Wang et al. 2010). In the same way, in the direction along the field line, the accelerating electric field is formed outside the null charge surface, where $\rho\sim{}0.5\rho_{GJ}$, and screened out inside the null charge surface where $|\rho|$ increases. It is still unclear what the distribution of the charge density and the electric field below the null charge surface are, and how the region with the two-layer structure connects to the region below the null charge surface. However, in this paper we discuss the radiation processes far from the null charge surface, so we can leave such questions to a future general study of the soft $\gamma$-ray and X-ray radiations from the spin-down powered pulsars.

The origin of these charged particles between the stellar surface and the null charge surface has been discussed by Takata et al. (2010). The photon-photon pair creation between the thermal X-ray photons and the GeV photons generates pairs near the null charge surface, and the magnetic field generates pairs near the stellar surface via magnetic pair creation. The magnetic field of the pulsar has two parts: the dipole field and the multi-pole field. The dipole field, which can be determined from the observed period $P$ and its derivative $\dot{P}$, dominates the magnetosphere of the neutron star, and determines the accelerating electric field together with the period of the pulsar $P$. The multi-pole field, which cannot be observed directly, dominates the region near the stellar surface and plays an important role in the gap closure. The charged particles under the null charge surface move inward to the stellar surface along the magnetic field lines, losing their energy by the curvature radiation. Because the accelerating electric field is weak, the lost energy cannot be compensated for by the electric field, and the energy of the incoming curvature photons emitted far from the null charge surface is around hundreds of MeV. If  these curvature photons are converted to pairs by the multi-pole field, part of the secondary pairs can obtain outgoing momentums from the multi-pole field and go outwards along the field lines to close the gap. These pairs may radiate synchrotron photons after being generated, but they can be neglected in this model, because 1) the photons may be converted to pairs by the multi-pole field, 2) they may be covered by the neutron star, and 3) the pairs spiral around the complex multi-pole field lines. These synchrotron photons cannot show the peak caused by the $caustic$ effect, which is due to the photons emitted from different places in the magnetosphere having parallel directions.

However, not all of the incoming curvature photons are converted to pairs by the multi-pole field; if the dipole field is strong enough the curvature photons can also be attenuated by the dipole field. If the pairs are generated by the dipole field, their synchrotron radiation has a chance to avoid entering the region dominated by the multi-pole field, and shows a significant pulse profile due to the dipole field's caustic effect.

\subsection{The magnetic field and the electric field}
We adopt the three dimensional rotating vacuum dipole field as the global magnetic field, which is given by
$
\vec{B}=\hat{r}[\hat{r}\cdot(\frac{3\mu}{r^3}+\frac{3\dot{\mu}}{cr^2}+\frac{\ddot{\mu}}{c^2r})]-(\frac{\mu}{r^3}+\frac{\dot{\mu}}{cr^2}+\frac{\ddot{\mu}}{c^2r})
$
(Cheng, Ruderman \& Zhang 2000), where $\mu=\mu(\hat{x}\sin{\alpha}\cos{{\Omega}t}
+\hat{y}\sin{\alpha}\sin{{\Omega}t}+\hat{z}\cos{\alpha})$ is the magnetic moment vector, $\hat{r}$ is the radial unit vector, and $\alpha$ is the inclination angle. We do not know the relationship between the surface multi-pole field and the global dipole field. For PSR B1509, a pulsar with a strong dipole field, we assume the multi-pole field dominates the region of $r<1.5R_s$, where $r$ is the distance to the center of the neutron star, and $R_s$ is the stellar radius.

The realistic magnetic field of the pulsar's magnetosphere is a compromise between the force free field and the rotating vacuum dipole field, which are two eigenstates of the magnetic field of the spin-down powered pulsar. Numerical results show that the two fields have distinct field structures, which provide different pulse profiles (Spitkovsky 2006; Bai \& Spitkovsky 2010). However, our model of the synchrotron radiation does not strongly require a realistic field structure, because the magnetic pair creation and the synchrotron radiation happen near the star, where the structures of the force free field and the vacuum dipole are close to each other.

For the accelerating electric field outside the null charge surface, we apply our three dimensional two-layer structure model (Wang et al. 2011). Although the outgoing curvature photons of PSR B1509 are missed by the viewing angle, for the integrity of the model we here repeat the equations:
\begin{equation}
E_{||}(x, z, \phi_p)\sim\frac{\Omega B}{cs}\left\{
\begin{array}{lcc}
-g_1z^2+C'_1z, \\ $for$~0\le{}z\le{}h_1 \\
g_2(z^2-h^2_2)+D'_1(z-h_2), \\ $for$~h_1<z\le{}h_2
\end{array},
\right.
\label{electric_3}
\end{equation}
where
\[
C'_1=-\frac{g_1h_1(h_1-2h_2)+g_2(h_1-h_2)^2}{h_2},
\]
\[
D'_1=-\frac{g_2h_2^2+(g_1+g_2)h_1^2}{h_2}.
\]
The $x$, $z$ and $\phi_p$ represent coordinates along the magnetic field line,
the height measured from the last-open field line, and the azimuthal direction. $h_1$ and $h_2$ represent the thickness of the main-accretion region and the total gap thickness, respectively. The $g=(\rho-\rho_{GJ})/\rho_{GJ}$ is the deviation of the charge density from the Goldreich-Julian charge density $\rho_{GJ}$ and in the unit of $\rho_{GJ}$:
\begin{equation}
g(z, \phi_p)=\left\{
   \begin{array}{ccc}
            -g_1(\phi_p), &$if$& 0\leq{}z\leq{}h_1\\
            g_2(\phi_p), &$if$& h_1<z\leq{}h_2
   \end{array}.\right.
\label{cdensity}
\end{equation}
where $g_1>0$, $g_2>0$ and they satisfy $(h_2/h_1)^2=1+g_1/g_2$. The $h_2$ and the polar radius $r_p$ satisfy the relationship, $f=h_2(R_s)/r_p$.

We do not consider the distributions of the three parameters, $1-g_1$, $h_1/h_2$ and $f$, in the azimuthal direction in the gap, which are found from the energy dependent light curves of the Vela pulsar (Wang et al. 2011). This is because the soft $\gamma$-ray photons of PSR B1509 are not the outgoing curvature photons. We use $1-g_1=0.05$, $h_1/h_2=0.92$, and $f=0.35$.

Under the null charge surface, there is so far no picture about the distribution of the charge density and the corresponding reduced accelerating electric field as clear as the two-layer structure for the region above the null charge surface. The electric field is assumed to decrease quadratically along the field line to the inner boundary of the gap, where $E_{\parallel}=0$. We adopt the form of the electric field used in Tang et al. (2008) and Takata et al. (2010):
\begin{equation}
E_{\parallel}(r<r_{null})=\frac{(r/r_{in})^2-1}{(r_{null}/r_{in})^2-1}E_{\parallel,null},
\label{E_inside}
\end{equation}
where the $r_{in}$ and $r_{null}$ are the distance of the inner boundary of the gap and the null charge surface from the star, respectively, and the $E_{\parallel,null}$ is the electric field at the null charge surface, which is given by Equation~(\ref{electric_3}). For PSR B1509, which has a period of 150ms, the position of the inner boundary is not important here, because the energy of the curvature photons emitted by the incoming particles are not determined by the weak electric field in the large magnetosphere, except in the region close to the null charge surface. We set the inner boundary as $r_{in}=20R_s$.

The electric field defined above can still accelerate the particles near the null charge surface to emit curvature photons even up to $>$1GeV. These particles go towards the stellar surface along the field lines and lose their energy through emitting curvature photons, whose energy drop down from GeV to hundreds of MeV. In order to know the number of these particles at any place between the stellar surface and the null charge surface, we set the number density of the incoming particles as,
\begin{equation}
n(r<r_{null})=\eta\frac{\Omega{}B(r)}{2\pi{}ec},
\label{rho_inside}
\end{equation}
which is of the same order as the local Goldreich-Julian charge density, and  $\eta$ is a fitting parameter. Because of the conservation of the magnetic flux, the flux of these incoming particles is a constant along the field line, from which we can calculate their curvature radiation. Along one field line,  $\eta$ should be continuous at the null charge surface, because the outgoing and incoming particles are generated from the pair creations occurring around the null charge surface. If we define the charge density above the null charge surface as positive, it nearly equals the density of the outgoing positrons, which can be described as $\rho(r>r_{null})\sim{}\rho_{+}$. On the other hand, under the null charge surface, because of the pair creation, the charge density is the combination of the density of positrons and electrons, $\rho(r\le{}r_{null})=\rho_{-}+\rho_{+}$, and $n_{-}(\vec{r})>|\rho(\vec{r})/e|$. Wang et al. (2010) show that the $\eta$ of the outgoing particles in the outer gap of the $\gamma$-ray pulsar is around 0.5, therefore, we chose $\eta=0.5$ for the incoming particles.

\subsection{The outgoing and incoming particles}
The evolution of the Lorentz factor of a particle moving along the magnetic field can be described as
\begin{equation}
mc^2\frac{d\gamma}{dt}=-\frac{2}{3}\gamma^4(\vec{r})\frac{e^2c}{s^2(\vec{r})}+eE_{\parallel}(\vec{r})c,
\label{equ_gamma}
\end{equation}
where $s(\vec{r})$ is the curvature radius of the magnetic field line at $\vec{r}$, the first term on the right side is the power of the curvature radiation, and the second term is the acceleration of the electric field.
From the null charge surface to the light cylinder, where the accelerating electric field $E_{\parallel}$ is strong, the outgoing particle can reach an equilibrium state such that it radiates the energy it obtains from the electric field as curvature radiation, which is described as $eE_{\parallel}(\vec{r})c=2e^2c\gamma^4(\vec{r})/3s^2(\vec{r})$, by requiring $d\gamma/dt=0$. The Lorentz factor of the outgoing particle in the equilibrium state is thus given by
\begin{equation}
\gamma_{equ}(\vec{r})=(\frac{3}{2}\frac{s^2(\vec{r})}{e}E_{\parallel}(\vec{r}))^{1/4},
\end{equation}
and the corresponding characteristic energy of the emitted curvature photon is given by
\begin{equation}
E_{equ}(\vec{r})=\frac{3}{2}\frac{\hbar{}c\gamma_{equ}^3(\vec{r})}{s(\vec{r})}.
\label{E_equ}
\end{equation}

For a particle moving from the null charge surface to the stellar surface, its Lorentz factor is calculated by solving Equation~(\ref{equ_gamma}) using the Runge-Kutta method.
Because the electric field in this region is much reduced, the energy loss by the curvature radiation of the particle cannot be compensated for by the potential drop of the gap. This situation has been discussed in Takata et al. (2010). When the Lorentz factor of the particle drops low enough, the curvature energy loss time scale becomes comparable to the time scale of the particle's movement to the stellar surface. There is a minimum energy $\sim 100$MeV for the curvature photon of a particle that is purely losing energy, and this energy does not depend on any pulsar parameters nor the curvature radius of the local magnetic field. As will be shown below, the energy of the synchrotron radiation is contributed by incoming particles between $20R_s$ and $50R_s$. Our numerical calculation shows that the characteristic energy of the emitted curvature photon $E_{in}(\vec{r})$ in this region spans from 200MeV to 400MeV.

\section{The pulse profile of the synchrotron radiation with a significant pitch angle}
Magnetic pair creation favours larger values of $\theta_p$, the pitch angle between the photon's propagation direction and the magnetic field. No magnetic pair creation will happen, no matter how strong the magnetic field is, if the photon's direction is parallel to the magnetic field line, which means $B_{\parallel}=0$. Conservation of momentum requires that the secondary pair generated by magnetic pair creation makes the same angle to the magnetic field  as that of the original curvature photon, while the opening angle of the pair  resulting from pair creation can be neglected compared with this pitch angle. The condition for magnetic pair creation cannot be satisfied at the birth place of the curvature photon, where the photon moves tangentially to the magnetic field line. The curvature photon should move along its path so as to encounter enough perpendicular magnetic field, and the secondary pair can therefore obtain a significant pitch angle from the pair creation, which cannot be neglected in the calculation of the pulse profile.

The equations of the motion of the particle in the acceleration region are (Harding, Usov \& Muslimov 2005; Hirotani 2006; Takata, Chang \& Shibata 2008):
\begin{equation}
\frac{dP_{\parallel}}{dt}=eE_{\parallel}-\frac{2e^2\gamma^2B^2\sin^2\theta_p}{3m^2c^4}\cos\theta_p,
\label{P_pal}
\end{equation}
and
\begin{equation}
\frac{dP_{\perp}}{dt}=\frac{c}{2B}\frac{dB}{ds}P_{\perp}-\frac{2e^2\gamma^2B^2\sin^2\theta_p}{3m^2c^4}\sin\theta_p.
\label{P_per}
\end{equation}
If the pairs are generated in the acceleration region, according to Equation~(\ref{P_pal}), the electric field's acceleration can reduce their pitch angles. On the other hand, if the pairs are generated in the region without strong accelerating electric field, and we neglect the first term of the right hand side of Equation~(\ref{P_per}), which represents the adiabatic change along the dipole field line, the motions of the secondary pairs can be described by,
\begin{equation}
\frac{dP_{\parallel}}{dt}=-\frac{2e^2\gamma^2B^2\sin^2\theta_p}{3m^2c^4}\cos\theta_p,
\label{dppa}
\end{equation}
and
\begin{equation}
\frac{dP_{\perp}}{dt}=-\frac{2e^2\gamma^2B^2\sin^2\theta_p}{3m^2c^4}\sin\theta_p,
\label{dppe}
\end{equation}
which indicate that the pitch angle $\theta_p$ stays constant while the particle loses its energy by emitting synchrotron photons.

In adopting our previous three dimensional studies of the radiation of pulsars, we introduce a factor $a$ to represent the magnetic field lines in the gap and take $a=1$ for the last-open field lines. We trace the field lines to determine the coordinate values  $[X_0 (\phi_p), Y_0(\phi_p), Z_0(\phi_p)]$ of the last-open field lines at the stellar surface. The coordinate values of an arbitrary layer of the magnetic field lines with a given ``$a$'', $[X'_0 (\phi_p), Y'_0(\phi_p), Z'_0(\phi_p)]$, can be determined by using $X'_0(\phi_p)=aX_0(\phi_p)$, $Y'_0(\phi_p)=aY_0(\phi_p)$ and $Z'_0(\phi_p)=(R^2_s-a^2X'^2_0-a^2Y'^2_0)^{1/2}$. The relation between the $z$ and the $a$ is approximated as
\begin{equation}
z(x,\phi_p)=\frac{1-a}{1-a_{min}}h_2(x,\phi_p),
\end{equation}
where $a_{min}$ corresponds to the upper boundary of the gap. In this paper, the dimensionless thickness $a_{min}$ is independent of $\phi_p$ and chosen as $a_{min}=0.925$, which corresponds to the fractional size of the gap $f$ we choose.

We divide the region of $0.925\le{}a\le{}1$ into 30 layers, which are represented by $a_i$ ($i=1, ..., 30$). We use 180 field lines to divide each layer into 180 tubes. Because of the inclination of the magnetic axis, some tubes' null charge surfaces are too close to the light cylinder to form acceleration regions. Therefore, in these 180$\times$30 tubes, we consider the radiations from the ones with $\phi_p\in{}[-70^{\circ}, 100^{\circ}]$, where the $\phi_p$ is the polar angle of the field line on the stellar surface. $\phi_p=0^{\circ}$ is at the right-hand side of the north pole's magnetic axis in the plane including the rotation axis and the magnetic axis, where the rotation axis is at the left-hand side of the magnetic axis of the north pole.

We trace the field lines in the gap, from the stellar surface to the light cylinder. At each point on the field line between the stellar surface and null charge surface, we follow these steps:

\begin{enumerate}
\item[1] We calculate the direction of the curvature photon $\vec{v}_{cur}$ (Takata, Chang \& Cheng 2007):
\begin{equation}
   \vec{v}_{cur}=\lambda{}v_p\vec{B}/B + \vec{\Omega}\times\vec{r},
   \label{direct_cur}
\end{equation}
where the first term represents the motion along the magnetic field line, $v_p$ is calculated from the condition that $|\vec{v}_{cur}|=c$, and the second term is the drift motion. If the magnetic field line is going outwards from the stellar surface, the outgoing direction is represented by $\lambda=1$, and the incoming direction is represented by $\lambda=-1$. If the magnetic field line is going towards the stellar surface, the outgoing and incoming directions are represented by $\lambda=-1$ and $\lambda=1$, respectively.
\\
\item[2] We trace the $\vec{v}_{cur}$ of the curvature photon emitted by the incoming particle to find the location of the pair creation. The thermal X-ray photons of typical temperature of a neutron star cannot convert the incoming curvature photons with hundreds of MeV into pairs. However, we cannot rule out the possibility that the keV soft synchrotron photons emitted from the gap of another pole can enter the gap and collide with the incoming curvature photons to become pairs, therefore we integrate both of the optical depths of the magnetic pair creation and photon-photon pair creation along the path of the curvature photon.

The optical depth of magnetic pair creation for a curvature photon with the energy $E_{\gamma}$ is
\begin{equation}
\tau_B(E_{\gamma})=\int{}l^{-1}(\vec{r}, E_{\gamma})dx,
\label{tauB}
\end{equation}
where $l(\vec{r}, E_{\gamma})$ is the mean free path of the curvature photon of $E_{\gamma}$ at position $\vec{r}$, which is given by (Erber 1966)
\begin{equation}
\begin{array}{l}
            l(\vec{r}, E_{\gamma})=\frac{4.4}{e^2/\hbar{}c}\frac{\hbar}{mc}\frac{B_q}{B_{\perp}(\vec{r})}\exp(\frac{4}{3\chi(\vec{r}, E_{\gamma})}),\\
            \chi(\vec{r}, E_{\gamma})\equiv\frac{E_{\gamma}}{2mc^2}\frac{B_{\perp}(\vec{r})}{B_q},
   \end{array}
\label{tauB2}
\end{equation}
where $B_q=m^2c^3/e\hbar=4.4\times{}10^{13}$G, $B_{\perp}=B\sin\theta_p$, and $\chi\ll{}1$ is required. Here the rotation of the magnetic field is considered, which in fact can be neglected for PSR B1509, because of its period of 150ms. For the photon emitted at $t=0$, the magnetic field at its position $\vec{r}$ when $t=t'$ is
\begin{equation}
\vec{B}(\phi, t')=M\vec{B}(\phi-\Omega{}t',0),
\label{rotate_B}
\end{equation}
where the $\phi$ is the position angle of the photon at $t=t'$, $M$ is a rotation matrix that rotates the magnetic field for $\Omega{}t'$ in the direction of position angle, and the position of the photon at $t'$ and its birth place $\vec{r}_0$ satisfy $\vec{r}-\vec{r}_0=t'\vec{v}_{cur}$.

The optical depth of photon-photon pair creation is
\begin{equation}
\tau_p(\vec{r}, E_{\gamma})=\int\int^{\epsilon_1}_{\epsilon_2}n_{soft}(\epsilon, \vec{r})\sigma(E_{\gamma}, \epsilon)d\epsilon{}dx,
\label{taup}
\end{equation}
where the cross section for photon-photon pair creation between the curvature photon with $E_{\gamma}$ and the soft photon with $\epsilon$, $\sigma(E_{\gamma}, \epsilon)$ is given by
\begin{equation}
\begin{array}{l}
\sigma(E_{\gamma}, \epsilon)=\frac{3}{16}\sigma_T(1-v^2)[(3-v^4)\ln(\frac{1+v}{1-v})-2v(2-v^2)],\\
v=\sqrt{1-(m_ec^2)^2/E_{\gamma}\epsilon},
\end{array}
\end{equation}
where $\sigma_T$ is the Thomson cross section. We assume that the angle between the directions of the two photons is 90$^{\circ}$ on average. The number density of the soft photons, $n_{soft}(\epsilon, \vec{r})$ is estimated as,
\begin{equation}
n_{soft}(\epsilon, \vec{r})=0.1\frac{F_{obs}(\epsilon)}{\epsilon^2c}\frac{d^2}{r^2},
\end{equation}
where the $F_{obs}$ is the observed flux of soft photon radiation in units of MeV/cm$^2$/s, $r$ is the distance to the center of the star, and $d$ is the distance of the pulsar. We use the observed data of 2$-$20 keV, the energy range of a photon that can convert photons of hundreds of MeV to a pair, provided by Ginga (Kawai et al. 1992). The value of $d$ is chosen as 4 kpc (Cordes \& Lazio 2002). Not all of the observed soft photons can enter the gap of another pole and have collisions under large angles with the curvature photons, therefore, instead of making a simulation to trace the paths of these soft photons, we assume $10\%$ of them join the photon-photon pair creation. However, as will be shown in the next section (Figure~\ref{Abs_spectrum}), the pairs emitting synchrotron photons are mainly generated by magnetic pair creation, and the photon-photon pair creation's contribution is negligible.
\\
\item[3] We integrate the two optical depths given above along $\vec{v}_{cur}$, and calculate the directions of the synchrotron photons, when $\Delta{\tau_B}(10 $MeV$\le{}E_{\gamma}\le{}1.5$GeV$)\ge{}0.1$ or $\Delta{\tau_p}(10 $MeV$\le{}E_{\gamma}\le{}1.5$GeV$)\ge{}0.1$. These conditions mean that $1-e^{-\Delta\tau_B(E_{\gamma})}\ge0.095$ or $1-e^{-\Delta\tau_p(E_{\gamma})}\ge0.095$ of the curvature photons of $E_{\gamma}$ at $\vec{r}$ become pairs. The $\vec{v}_{syn}$ in the hollow cone of the synchrotron radiation can be expressed as,
\begin{equation}
   \vec{v}_{syn}=\lambda{}v_p\vec{v}'_{syn}/|\vec{v}'_{syn}| + \vec{\Omega}\times\vec{r},
\end{equation}
where $v_p$ is calculated from the condition $|\vec{v}_{syn}|=c$ and $\vec{v}'_{syn}$ is given by,
\begin{equation}
  \begin{array}{ccc}
       v'_{syn,x}= \hat{B}_x + \tan{\theta_p}(u_x\cos{}T + v_x\sin{}T)\\
       v'_{syn,y}= \hat{B}_y + \tan{\theta_p}(u_y\cos{}T + v_y\sin{}T)\\
       v'_{syn,z}= \hat{B}_z + \tan{\theta_p}(u_z\cos{}T + v_z\sin{}T)\\
  \end{array}.
  \label{direction_cone}
\end{equation}

In the equations above, $\hat{B}$ is the normalized local magnetic field, $\vec{u}=(\frac{B_y}{(B_x^2+B_y^2)^{1/2}}, -\frac{B_x}{(B_x^2+B_y^2)^{1/2}},0)$ and $\vec{v}=\frac{\vec{B}\times\vec{u}}{|\vec{B}\times\vec{u}|}$ are two orthogonal vectors perpendicular to the magnetic field. $\theta_p$ is the pitch angle of the pair, which is the same as that of the curvature photon. For  photons of hundreds of MeV, the opening angle of the pair obtained from the pair creation is much smaller than the pitch angle of the curvature photon and therefore is negligible. If $\theta_p\le{}90^{\circ}$, then $\lambda=1$; and if $\theta_p>90^{\circ}$, $\lambda=-1$. $T\in[0, 2\pi)$ represents a direction along the cone, we trace the $\vec{v}_{syn}$ of each $T$ and remove the photons that will enter the region dominated by the multi-pole field. The velocities of the photons will be used to calculate their pulse phases, therefore they should be measured at the same time, and the $\vec{v}_{syn}$ should be rotated $-\Omega{}t'$ in the direction of position angle.
\\
\item[4] Having the direction of the photon, $\vec{v}_{syn}$ or $\vec{v}_{cur}$, the viewing angle $\zeta$, which is the angle to the rotation axis of the emission's direction,
and the pulse phase $\psi$ of the synchrotron or curvature photon can be obtained from (Yadigaroglu 1997)
\begin{equation}
\left\{
   \begin{array}{ccc}
             \cos{\zeta}=v_{z}/|\vec{v}|\\
             \psi=-\cos^{-1}(v_{x}/v_{xy})-\vec{r}\cdot\hat{v}/R_L
   \end{array}.\right.
\end{equation}

In the skymap (e.g. Figure~\ref{two_lines}), where the X-axis is the pulse phase $\psi$ and the Y-axis is the viewing angle $\zeta$, the highly beamed curvature photons emitted from $\vec{r}$ are represented by a point $(\psi, \zeta)$; on the other hand, the synchrotron photons emitted from $\vec{r}$ are represented by a line, as the $T$ of Equation~(\ref{direction_cone}) increases from 0 to $2\pi$. In practice, the increment of $T$, the $\Delta{T}$, should make $|\Delta{\zeta}|$ and $|\Delta{\psi}|$ stay nearly constant. A visible synchrotron photon satisfies $|\zeta-\beta|<0.5^{\circ}$ and its $\vec{v}_{syn}$ should not point to the region dominated by the multi-pole field.
\end{enumerate}

\begin{figure}[t]
\plotone{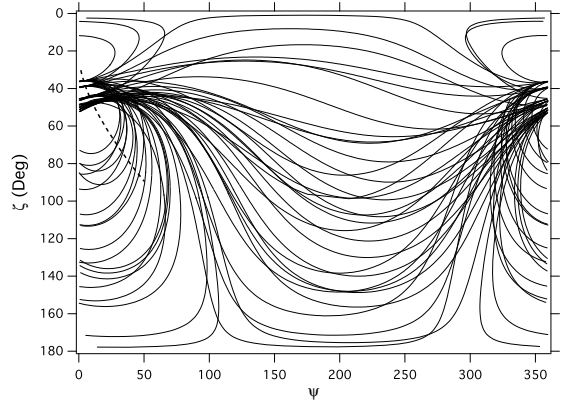}
\caption{The skymap of the viewing angle $\zeta$ and the pulse phase $\psi$ of the synchrotron radiation (solid line) and the incoming curvature radiation (dashed line) originating from one magnetic field line, where the inclination angle is $\alpha=20^{\circ}$. The incoming curvature photons, which are represented by the points on the dashed line, are emitted by the particles moving from the null charge surface to the stellar surface of one magnetic field line. The synchrotron photons are emitted from the place, where $\Delta{\tau_B}(10 $MeV$\le{}E\le{}1.5$GeV$)\ge{}0.1$ or $\Delta{\tau_p}(10 $MeV$\le{}E\le{}1.5$GeV$)\ge{}0.1$. Along the directions of the curvature photons, there are many places where these conditions are satisfied. The hollow cones of the synchrotron radiation are represented by the solid lines.}
\label{two_lines}
\end{figure}

Figure~\ref{two_lines} shows the differences between the $(\psi, \zeta)$ of curvature radiation and those of the synchrotron radiation with a significant pitch angle. In this figure, the curvature photons (dashed line) and the synchrotron photons (solid lines) originate from the region between the null charge surface and the stellar surface of only one magnetic field line. In the direction of the curvature photons emitted from $\vec{r}$, which is represented by a point on the dashed line, there are about $\tau/\Delta{}\tau$ hollow cones with different pitch angles, where $\tau$ is the optical depth of the pair creation integrated from $\vec{r}$ to the light cylinder along $\vec{v}_{cur}$, and $\Delta{}\tau=0.1$. Each hollow cone is represented by a solid line in this skymap.
Because the hollow cone of the synchrotron radiation of the secondary pair contains the direction of the original curvature photon, the lines of the synchrotron radiation in the skymap, which have the same birth place of the original curvature radiation, intersect at the $(\psi, \zeta)$ of the original curvature radiation. A large step length is used to trace this field line, hence there are nearly five points of the curvature radiations, where the lines of synchrotron radiation intersect.

\begin{figure}[t]
\plotone{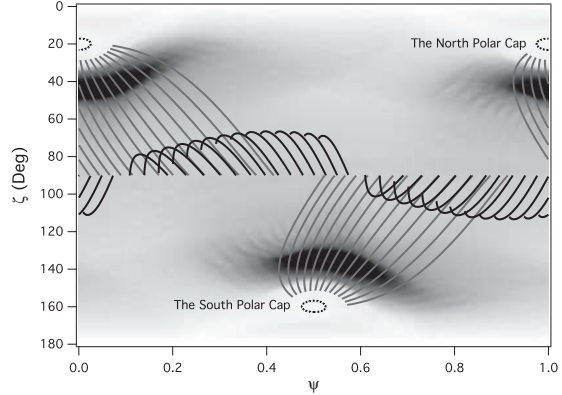}
\caption{The skymap of the viewing angle $\zeta$ and the pulse phase $\psi$ of the outgoing curvature photons (black line) emitted from the null charge surface to the light cylinder,  the incoming curvature photons (gray line) emitted from the null charge surface to the stellar surface, and the synchrotron photons (darkness) emitted from the place where $\Delta{\tau_B}(10 $MeV$\le{}E\le{}1.5$GeV$)\ge{}0.1$ or $\Delta{\tau_p}(10 $MeV$\le{}E\le{}1.5$GeV$)\ge{}0.1$. The curvature photons are from the lines of $a=0.95$, but the synchrotron photons' $\zeta$ and $\psi$ are calculated by tracing the field lines within $\phi_p\in{}[-70^{\circ}, 100^{\circ}]$ of $a=0.925 - 1.0$. The inclination angle is $\alpha=20^{\circ}$.}
\label{lc_colmap}
\end{figure}

\begin{figure}
\plotone{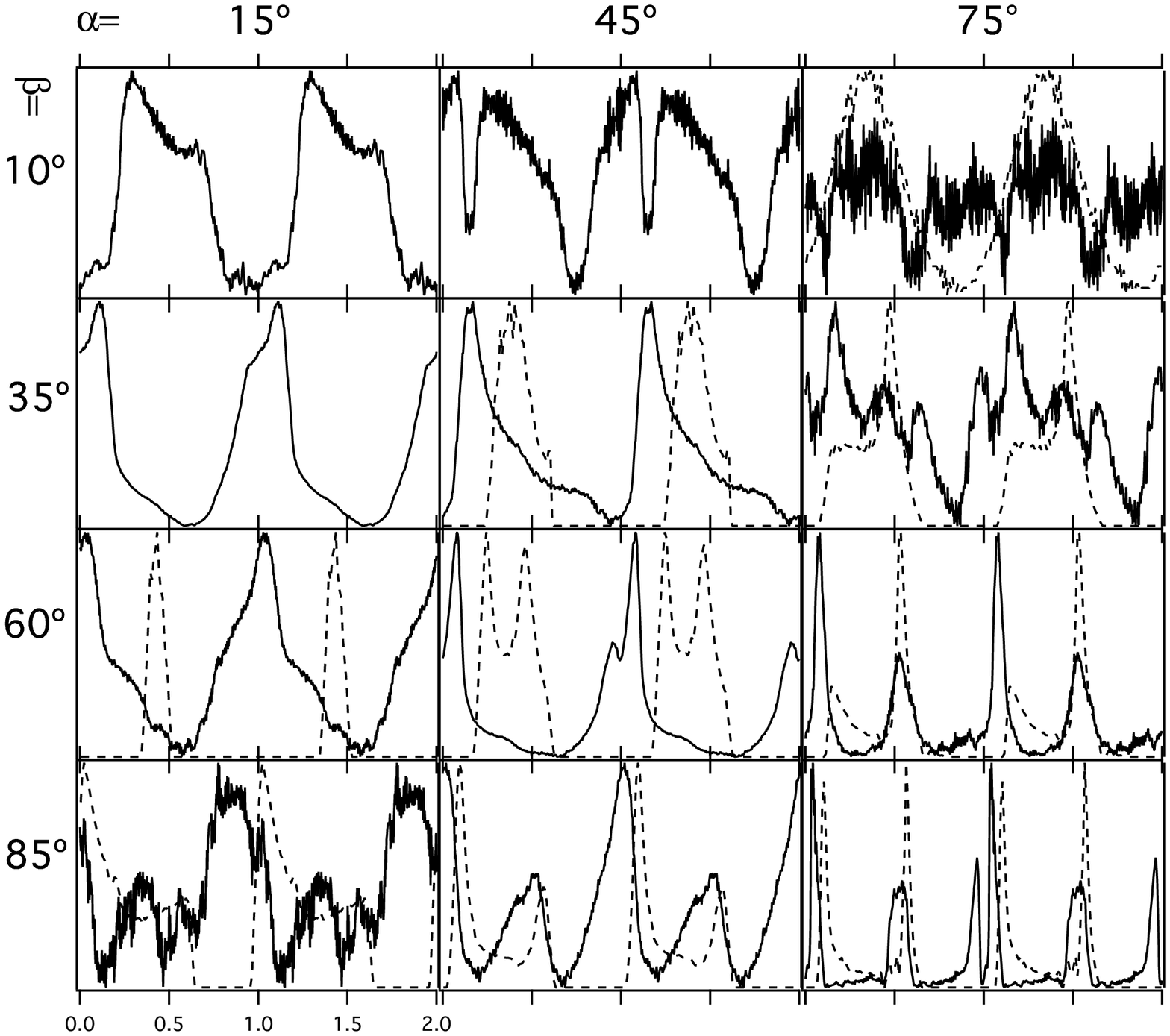}
\caption{The light curves of two rotation periods of the synchrotron radiation (solid lines) ignoring the attenuation of the original incoming curvature radiation, and the light curves of the outgoing curvature radiation (dashed lines), of different inclination angles and viewing angles. The two types of radiation have different scales on y-axis.}
\label{diff_a_b}
\end{figure}

We use a smaller step length to trace the field lines within $\phi_p\in{}[-70^{\circ}, 100^{\circ}]$ of $a=0.925-1.0$. The synchrotron radiation of the secondary pairs is shown by the color map in Figure~\ref{lc_colmap}, where the inclination angle $\alpha=20^{\circ}$, the black lines and the gray lines are the outgoing curvature radiation and the incoming curvature radiation emitted from the field lines of $a=0.95$, respectively. The outgoing curvature photons can be missed by the viewing angle $<60^{\circ}$ under $\alpha=20^{\circ}$. Furthermore, if the viewing angle is smaller than the inclination angle, even the survival incoming curvature photons can be missed. Therefore, small inclination angle and viewing angle may be the reason for the disappearance of the GeV photons. In this figure, the dot circles are the pulse phases and viewing angles of the two polar caps. It is widely accepted that the radio emission comes from the polar caps for most of the $\gamma$-ray spin-down powered pulsars, and the pulse phase of the observed radio emission is that of one of the two polar caps, which is set to be 0 here. Figure~\ref{diff_a_b} lists the light curves of two rotation periods of the outgoing curvature radiation (dashed lines) and the synchrotron radiation (solid lines) of the pairs generated by the magnetic pair creation. The light curves of the two kinds of radiation have different scales in the y-axis. As the inclination angle and the viewing angle decrease, the pulse profile of the synchrotron radiation gets closer to the observed one of PSR B1509, i.e. a wide single peak. So, a small inclination and a small viewing angle are required both by the pulse profile of the soft $\gamma$-ray radiation and the absence of the GeV photons.


The light curve of the synchrotron radiation read from the color map, however, is close to but not the final result, because it is not energy dependent and the calculation of the color map does not consider the local emissivity or the attenuation of the original incoming curvature photons, which can change the shape of the light curve. In order to get the energy dependent light curves, we have to calculate the phase resolved spectra.

\section{The spectrum and the energy dependent light curves of the X-ray and soft $\gamma$-ray emissions}
\label{cal_spectrum}
The secondary pairs are generated in the region where the magnetic field can satisfy $\frac{B^2}{\epsilon_Xn_X}\gg{}1$, where $\epsilon_Xn_X$ is the energy density of the soft photons that can collide with the pairs. This means that the pairs lose their energy mainly via synchrotron radiation, and Inverse Compton scattering can be neglected. Because the observed spectra are non-thermal, we neglect the blackbody radiation from the stellar surface.

We follow the four steps described above to find the positions where the radiations' directions satisfy $|\zeta-\beta|<0.5^{\circ}$. The photon spectrum of the synchrotron radiation emitted from $\vec{r}$ is given by
\begin{equation}
\frac{d^2N_{syn}(E, \vec{r})}{dEdt}=\frac{\sqrt{3}e^3B(\vec{r})\sin\theta_p}{2\pi{}m_ec^2\hbar{}E}\int^{E_{max}}_{E_{min}}\frac{dN_e(\vec{r})}{dE_e}F(\kappa)dE_e,
\label{spec_syn}
\end{equation}
where the $E_{max}$ and $E_{min}$ are the maximum and minimum energy of the secondary positron or electron, respectively. The energy of the pair should be in the range of $mc^2<E_e\le{}E_{cur}/2$, where the $E_{cur}$ is the energy of the original curvature photon. $F(\kappa)=\int^{\infty}_{\kappa}K_{5/3}(\xi)d\xi$, where $K_{5/3}$ is the modified Bessel functions of order 5/3, and $\kappa=E/E_{syn}(\vec{r}, E_e)$, $E_{syn}(\vec{r}, E_e)$ is the critical synchrotron photon energy, which is given by
\begin{equation}
E_{syn}(\vec{r}, E_e)=\frac{3\hbar{}eB(\vec{r})\sin\theta_p(\vec{r}){}E_e^2}{2m_e^3c^5}.
\label{Esyn_typ}
\end{equation}
The energy distribution of the secondary pairs at $\vec{r}$ is
\begin{equation}
\begin{array}{c}
\frac{dN_e(\vec{r}, \epsilon_e)}{dE_e}=\frac{1}{\dot{E_e}(\vec{r}, \epsilon_e)}\int^{E_{max}}_{\epsilon_e}\frac{d^2N_{cur}(\vec{r}_0, E_{\gamma}=2E')}{dE_{\gamma}dt}\\
e^{-\tau(E_{\gamma}, \vec{r})}(1-e^{-\Delta\tau(E_{\gamma}, \vec{r})})dE',
\end{array}
\end{equation}
where $\dot{E_e}(\vec{r}, E_e)=2e^4B^2(\vec{r})\sin^2{\theta_p(\vec{r})}E^2_e/3m^4_ec^7$ is the synchrotron energy-loss rate of the particle with energy $E_e$, $E_{max}$ is the maximum energy of the original curvature photons, and $\tau(E_{\gamma}, \vec{r})=\tau_B(E_{\gamma}, \vec{r})+\tau_p(E_{\gamma}, \vec{r})$ is the integrated optical depth from $\vec{r}_0$, the place where the original curvature photons are emitted, to $\vec{r}$. The definitions of the two optical depths are given by Equation~(\ref{tauB}) and Equation~(\ref{taup}). $\Delta\tau(E_{\gamma}, \vec{r})$ is the sum of the two optical depths between $\vec{r}$ and $\vec{r}+\Delta{\vec{r}}$, and $\Delta{}\vec{r}$ is the step we take. $(\vec{r}-\vec{r}_0), \Delta{}\vec{r}$ and $\vec{v}_{cur}$ are parallel.

The photon spectrum of the curvature radiation emitted from $\vec{r}_0$, is
\begin{equation}
\frac{d^2 N_{cur}(\vec{r}_0, E_{\gamma})}{d E_{\gamma} d t}=\frac{\Delta{N(\vec{r}_0)}
\sqrt{3}e^2\gamma}{2\pi{\hbar}s(\vec{r}_0)E_{\gamma}
}F(\kappa),
\label{spec_cur}
\end{equation}
where $s(\vec{r}_0)$ is the local curvature radius of the magnetic field line, $F(\kappa)$ is the same as that in Equation~(\ref{spec_syn}), but $\kappa=E_{\gamma}/E_{cur}(\vec{r}_0)$, while $E_{cur}$ and $\gamma$ are the characteristic energy of the radiated curvature photons and the corresponding Lorentz factor of the particle respectively. For incoming curvature radiation, $E_{cur}$ is obtained by solving Equation~(\ref{equ_gamma}); while for outgoing curvature radiation, $E_{cur}$ is given by Equation~(\ref{E_equ}). $\Delta{N}(\vec{r}_0)=\Delta{V}(\vec{r}_0)n(\vec{r}_0)$, where $n(\vec{r}_0)$ is the number density of the accelerated particle, and $\Delta{V}(\vec{r}_0)$ is the volume of the tube-like region at $\vec{r}_0$, which can be obtained from the conservation of the magnetic flux,
\begin{equation}
\Delta{V(\vec{r}_0)}=\frac{2{\pi}r_{p}h_2(R_{s})B_s\Delta{l}(\vec{r}_0)}{N_AN_BB(\vec{r}_0)},
\end{equation}
where $\Delta{l}$ is the step length used to trace the field line, $r_p$ is the polar radius of the gap, and $N_A\times{}N_B=180\times30$ is the number of the lines we use to divide the region of $a=0.925 - 1$. For the outgoing particles outside the null charge surface, the number density $n(\vec{r}_0, r_0>r_{null})=\Omega{}B(\vec{r}_0)[1-g(z)]/2\pi{}ec$; for the incoming particles inside the null charge surface, the number density is defined in Equation~(\ref{rho_inside}), $\eta$ is chosen as $\eta=0.5$, which equals that of the outgoing particles.

If the incoming curvature radiation is emitted far enough from the stellar surface, the survival radiation from the attenuation caused by the pair creation has the chance to be observed at certain viewing angles, given by
\begin{equation}
\frac{d^2 N_{cur, sur}(\vec{r}_0, E_{\gamma})}{d E_{\gamma} d t}=\frac{d^2 N_{cur}(\vec{r}_0, E_{\gamma})}{d E_{\gamma} d t}\exp(-\tau(E_{\gamma}, \vec{r'})),
\end{equation}
the $\vec{r'}$ is the position where we stop tracing the path of the curvature radiation and integrating the two optical depths along $\vec{v}_{cur}$. Some of the synchrotron photons may penetrate to the region where the magnetic field can convert them into pairs. Therefore, we trace the trajectories of the synchrotron radiation with $|\zeta-\beta|<0.5^{\circ}$ along $\vec{v}_{syn}$ and integrate the optical depth of the magnetic pair creation. Here, photon-photon pair creation can be neglected. The survival synchrotron radiation is given by
\begin{equation}
\frac{d^2 N_{syn, sur}(\vec{r}, E)}{d E d t}=\frac{d^2 N_{syn}(\vec{r}, E)}{d E d t}\exp(-\tau_B(E_{\gamma}, \vec{r'})),
\end{equation}
where $\vec{r'}$ is the place where $\tau_B$ becomes constant. 

The total photon flux received at Earth is the sum of the survival synchrotron radiation, survival incoming curvature radiation and the outgoing curvature radiation,
\begin{equation}
\begin{array}{c}
F_{tot}(E)=\frac{1}{D^2}[ \sum_{i}\frac{1}{\Delta{\Omega_{syn}}(\vec{r}_{i})}\frac{d^2 N_{syn, sur}(E, \vec{r}_{i})}{d E d t} +\\
\sum_{j}\frac{1}{\Delta{\Omega_{cur}}(\vec{r}_{j})}\frac{d^2 N_{cur, in, sur}(E, \vec{r}_{j})}{d E d t} +\\
\sum_{k}\frac{1}{\Delta{\Omega_{cur}}(\vec{r}_{k})}\frac{d^2 N_{cur, out}(E, \vec{r}_{k})}{d E d t}
 ].
\end{array}
\end{equation}
For PSR B1509, the curvature radiation is outside the viewing angle. The distance of PSR B1509 is chosen as $D=4$kpc, which does not conflict with the H$_{I}$ absorption measurement: 5.2$\pm$1.4 kpc (Gaensler et al. 1999), and the dispersion measurement: 4.2$\pm$0.6 kpc (Cordes \& Lazio 2002). The local solid angle of the curvature radiation is given by
\begin{equation}
\begin{array}{l}
\Delta\Omega_{cur}(\vec{r})\simeq{}\pi{}\theta^2,\\
\end{array}
\end{equation}
where $\theta$ is chosen as $0.5^{\circ}$, according to the criterion of the visible radiation $|\zeta-\beta|<0.5^{\circ}$.
Because the secondary pairs are generated at places without an accelerating electric field, as shown in Equation~(\ref{dppa}) and Equation~(\ref{dppe}), their pitch angles are constant, and the solid angle of the synchrotron radiation emitted from $\vec{r}$ is that of a hollow cone, which is given by
\begin{equation}
\Delta{\Omega}_{syn}(\vec{r})=2\pi\int^{\theta_p+\frac{1}{2\gamma}}_{\theta_p-\frac{1}{2\gamma}}\sin\theta{}d\theta,
\end{equation}
where $\gamma$ is the Lorentz factor of the particle. $\Delta{\Omega}_{syn}(\vec{r})$ depends on the energy of the particle, as the particle loses its energy via synchrotron radiation, $1/\gamma$ becomes larger and the resultant $\Delta{\Omega}_{syn}$ becomes larger. This makes the observed flux smaller. However, at the same time, the invisible synchrotron radiation emitted from other places becomes visible under larger $1/\gamma$. We assume that the two effects cancel out each other, and use $\Delta{\Omega}_{syn}(\gamma=250$MeV$/m_ec^2)$.

By fitting the observed soft $\gamma$-ray spectrum and the energy dependent light curves, we find that the inclination angle and the viewing angle of PSR B1509 are 20$^{\circ}$ and 11$^{\circ}$, respectively, which are different from those of the fitting by Romani \& Yadigaroglu (1995) and the fitting by Zhang \& Cheng (2000), where the inclination angle is chosen as $\alpha=60^{\circ}$. The fitting of the observed polarization profile of the radio emission, by using the rotating-vector-model (Radhakrishnan \& Cooke 1969), can provide a constraint on the inclination angle. The fitting of the PSR B1509 data by Crawford et al. (2001) suggests that $\alpha<60^{\circ}$, which does not conflict with our result.

\begin{figure}
\plotone{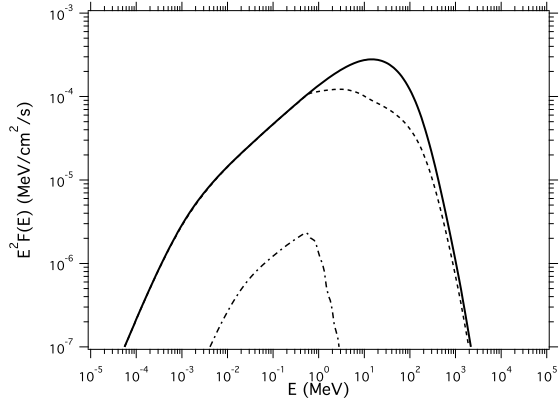}
\caption{The spectra of the synchrotron radiations of the secondary pairs generated by pair creations. The solid line is the one without considering the attenuation of the synchrotron photons by the magnetic pair creation. The dashed line is the spectrum of the survival synchrotron photons from magnetic pair creation. The dot-dashed line is the contribution of photon-photon pair creation, which considers the attenuation caused by the magnetic pair creation.}
\label{Abs_spectrum}
\end{figure}

Figure~\ref{Abs_spectrum} shows the spectra of $\alpha=20^{\circ}$ and $\beta=11^{\circ}$, where the outgoing curvature radiation and the possible survival incoming curvature radiation are missed by the line of sight. The dashed and solid lines are the cases with and without the attenuation of the synchrotron radiation of the secondary pair caused by the magnetic field, and show this process cannot be neglected. The reason for the low cut off energy of the spectrum of PSR B1509 is that the magnetic field converts some of the synchrotron photons into pairs. This figure also shows the contribution of the pairs generated by photon-photon pair creation (dot-dashed line), which confirms that the photon-photon pair creation is negligible and the pairs emitting synchrotron photons are mainly generated by the magnetic pair creation. Because these pairs are generated close to the stellar surface, where the soft photon density is high enough to allow photon-photon pair creation, their synchrotron radiations above 1MeV are strongly attenuated by the magnetic field.

\begin{figure}
\plotone{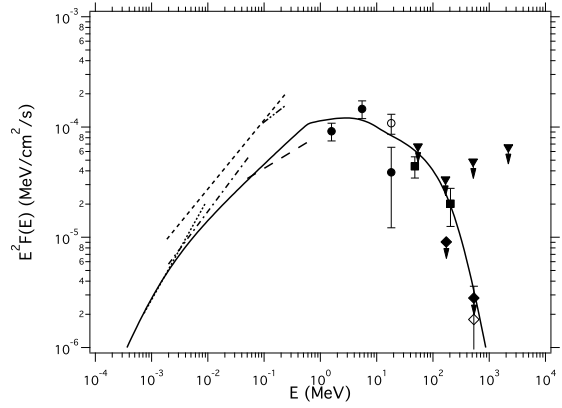}
\caption{The simulated phase averaged spectrum of PSR B1509, compared with the observed data. The triangles are the 2$\sigma$ upper limits for the total fluxes in the EGERT energy domain. The filled circles are the COMPTEL flux points derived from the excess counts in the 0.15$-$0.65 phase range, and the open circle is the 10$-$30MeV flux in the 0.15$-$0.65 phase range above the spatially determined background (Kuiper et al. 1999). The filled squares are the observed pulsed fluxes by AGILE in the 30$-$100MeV and 100$-$500MeV bands (Pilia et al. 2010). The filled diamonds are the 2$\sigma$ upper limits of $Fermi$ LAT derived with $gtlike$, and the open diamond is the integrated flux derived from the light curves observed by $Fermi$ (Abdo et al. 2010). The lines are the X-ray spectra observed by ASCA (dotted line) (Saito et al. 1997), Ginga (dot-dashed line) (Kawai et al. 1992), OSSE (long-dashed line) (Matz et al. 1994), Welcome (dot-dot-dashed line) (Gunji et al. 1994), and RXTE (short-dashed line) (Marsden et al. 1997).}
\label{PA_spectrum}
\end{figure}

Figure~\ref{PA_spectrum} compares the simulated phase averaged spectrum with the observed one. The simulated spectrum has different shapes above and below 1 MeV. Below 1MeV the spectrum is the same as that without the attenuation caused by the magnetic field; while above 1MeV the spectrum has an unusual shape with a low cut off energy. There is an estimation of the optical depth of the magnetic pair creation by Ruderman \& Sutherland (1975), which is, when $\chi$ in Equation~(\ref{tauB2}) satisfies
\begin{equation}
\chi^{-1}\approx{}15,
\label{appro_tauB}
\end{equation}
$\tau_B$ is unity. Here we define a parameter $E_{cri}(\vec{r})$, which is the maximum energy of the photon that can survive from the magnetic pair creation at the position $\vec{r}$ as
\begin{equation}
E_{cri}(\vec{r})=\frac{2mc^2}{15}\frac{B_q}{B_{\perp}(\vec{r})}.
\label{def_ecri}
\end{equation}
$E_{cri}=1$MeV corresponds to the maximum energy of synchrotron photons surviving from the magnetic pair creation. The maximum $B_{\perp}$ these photons encounter is about 3$\times{}10^{12}$G, so the spectrum with $E<1$MeV can maintain its shape of the synchrotron radiation, but the part with $E>1$MeV is changed by the magnetic field.

\begin{figure}
\plotone{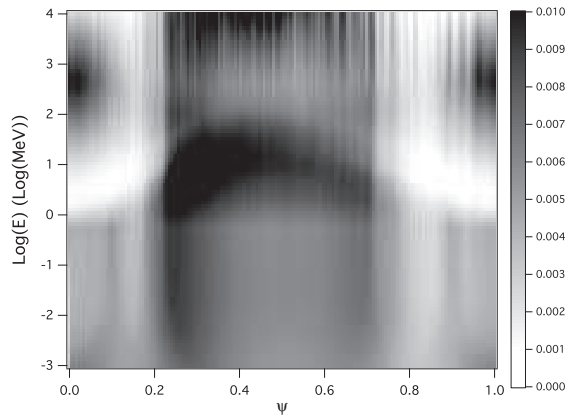}
\caption{The simulated energy dependent light curves in the pulse phase and energy plane. The darkness represents the percentage of the number of photons of a certain pulse phase interval, in the total number of photons of a certain energy range.}
\label{Colmap_LC}
\end{figure}

\begin{figure}
\plotone{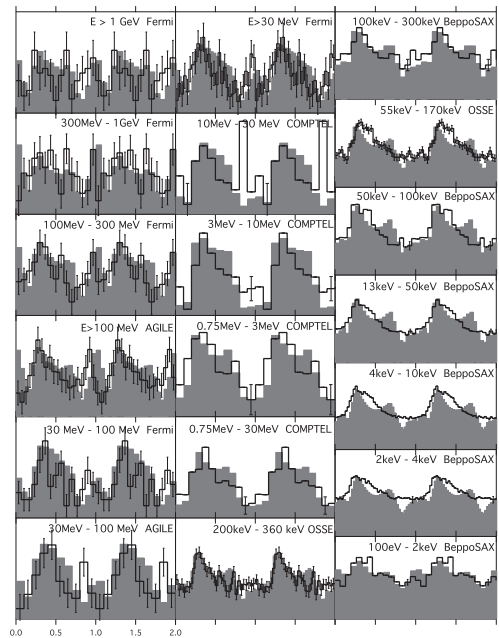}
\caption{The simulated energy dependent light curves (gray histograms), compared with the observed data (solid lines), which are provided by $Fermi$ LAT (Abdo et al. 2010), AGILE (Pilia et al. 2010), COMPTEL (Kuiper et al. 1999), OSSE (Matz et al. 1994) and BeppoSAX (Cusumano et al. 2001).}
\label{EDLCs}
\end{figure}

To obtain the energy dependent light curves, we bin the photons by their pulse phases and energy. The number of the photons with $E_1\le{}E\le{}E_2$
 measured at pulse phases between $\psi_1$ and $\psi_2$ is calculated from
\begin{equation}
N_{\gamma}(E_1, E_2, \psi_1, \psi_2)\propto\int^{E_2}_{E_1}{F_{tot}(E, \psi_1 \leq \psi \leq \psi_2)}dE.
\label{int_spec}
\end{equation}
As the secondary particle loses its energy via synchrotron radiation, its Lorentz factor $\gamma_e$ becomes smaller and the hollow cone can be observed for longer.
In order to take into account this fact, we redistribute the photons with  energy $E$ of $\psi$ into $\psi\pm{}\gamma_e^{-1}$, where $\gamma_e$ is the Lorentz factor of the pair emitting these photons with typical synchrotron energy $E$ under the local $B_{\perp}(\vec{r})$. 
Figure~\ref{Colmap_LC} shows the integrated energy dependent light curves in the plane of pulse phase and energy, where there are 100 bins in the axis of energy and 180 bins in the axis of pulse phase, which means that there are 100 energy dependent light curves integrated from 180 phase resolved spectra. The darkness represents the percentage of the number of the photons of a certain pulse phase interval, in the total number of photons of certain energy range. Below 10MeV, there is an asymmetric wide single pulse at $\psi\sim0.25$, and above 10MeV, another peak shows up at $\psi\sim{}0$. The explanation for such a phenomenon will be given in Section~\ref{Dis_2}. We compare the energy dependent light curves with the observed ones in Figure~\ref{EDLCs}.

\section{Discussion}
\subsection{The lost $\gamma$-ray photons}

The magnetosphere of PSR B1509, in fact, can also generate GeV $\gamma$-ray photons by accelerating the charged particles in the outer gap, just as in other $\gamma$-ray spin-down powered pulsars. However, these outgoing curvature photons are missed by our line of sight, which happens more easily under a smaller inclination angle. And because the viewing angle is smaller than the small inclination angle, even the survival incoming curvature photons are totally invisible. This requires us, together with the observed wide asymmetric single peak, to choose $\beta=11^{\circ}$ and $\alpha=20^{\circ}$, which does not conflict with the observed polarization of the radio emission.

\begin{figure}
\plotone{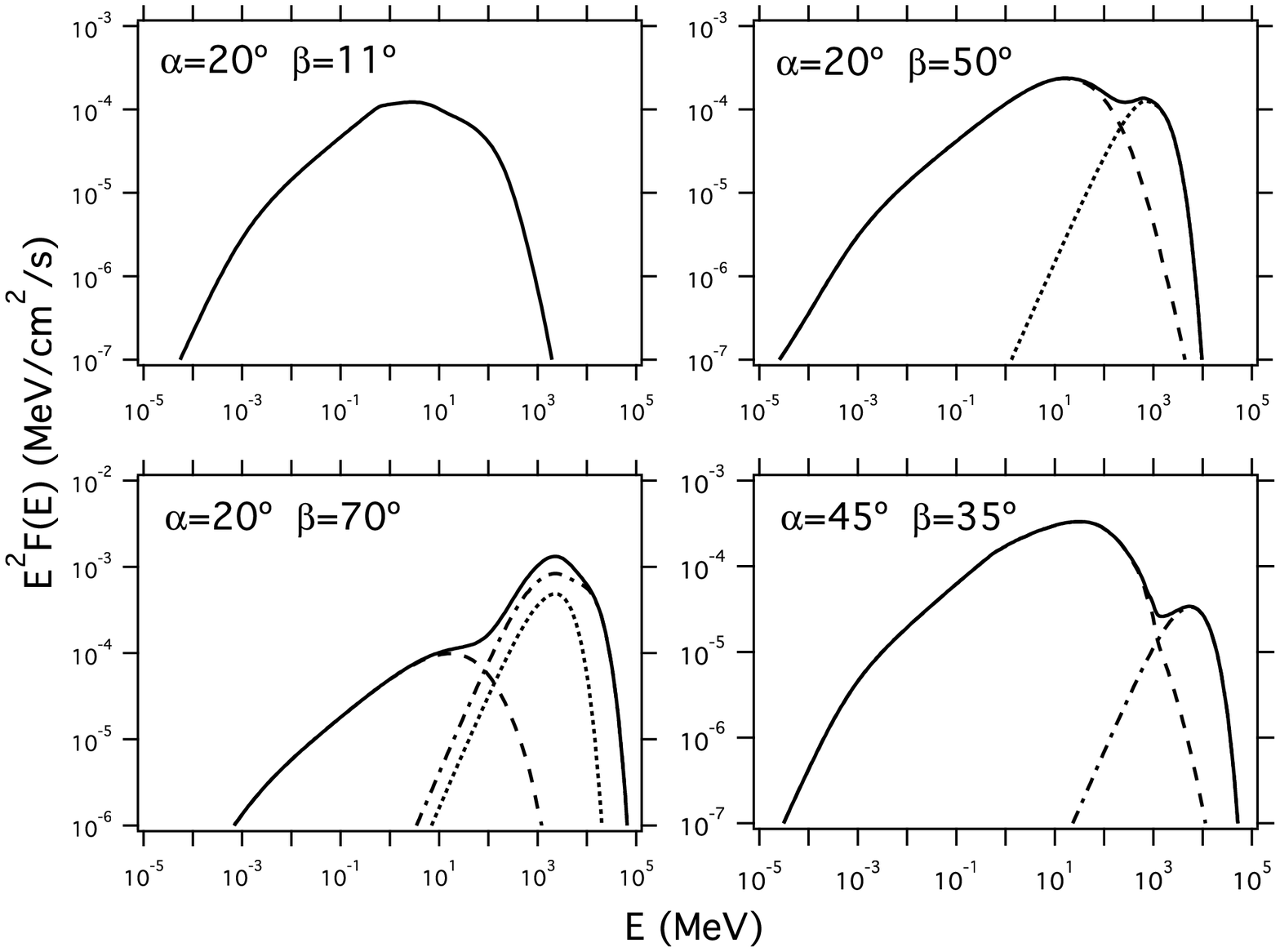}
\caption{The simulated phase averaged spectra of different inclination angles $\alpha$ and different viewing angles $\beta$. The dashed lines are the synchrotron radiations of the secondary pairs, the dot-dashed lines are the outgoing curvature radiations, the dotted lines are the survival incoming curvature radiations, and the solid lines are the total spectra.}
\label{S_Diff_ab}
\end{figure}

Figure~\ref{S_Diff_ab} shows the phase averaged spectra of different inclination angles and viewing angles. Under $\alpha=20^{\circ}$, as the viewing angle increases, the incoming and outgoing curvature radiations become visible. In a huge magnetosphere with a small inclination angle, the incoming curvature photons have a chance to escape from being converted into pairs by the magnetic field. Along one field line, the place of $\beta=70^{\circ}$ is closer to the null charge surface than that of $\beta=50^{\circ}$, therefore, the incoming curvature photons have higher energy. When $\alpha$ becomes larger, such as 45$^{\circ}$, the outgoing curvature radiation shows up earlier than the incoming one when $\beta$ increases. To obtain a clear picture of the connection between the two-layer structure and the region emitting incoming curvature photons besides the null charge surface, a general multiwavelength study of the high energy radiations of the pulsars is needed, where the different $\alpha$ and $\beta$ are considered.

\subsection{The second peak of $E>10$MeV}
\label{Dis_2}
\begin{figure}
\plotone{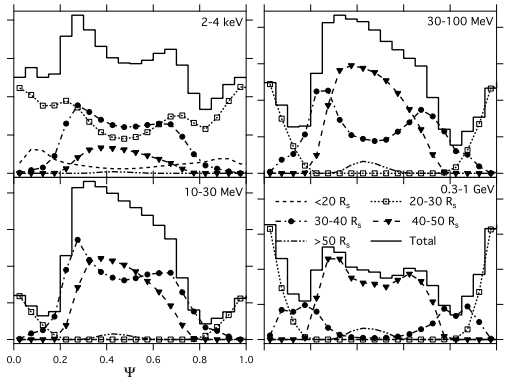}
\caption{The energy dependent light curves of different birth places of the original curvature photons.}
\label{LC_section}
\end{figure}

\begin{figure}
\plotone{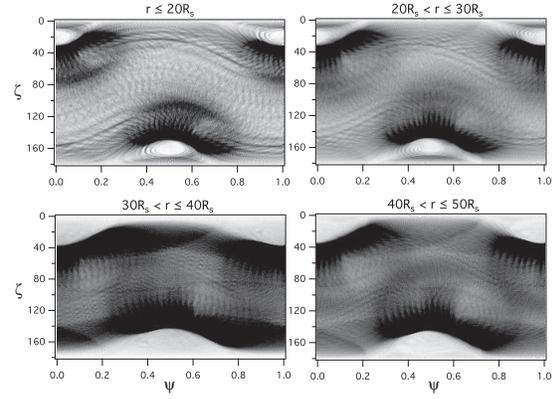}
\caption{The skymaps of the viewing angle $\zeta$ and the pulse phase $\psi$ of the synchrotron photons emitted from the place where $\Delta{\tau_B}(10 $MeV$\le{}E\le{}1.5$GeV$)\ge{}0.1$ or $\Delta{\tau_p}(10 $MeV$\le{}E\le{}1.5$GeV$)\ge{}0.1$. $r$ is the distance of the place where the original curvature photons are emitted.}
\label{Skymap_section}
\end{figure}

\begin{figure}
\plotone{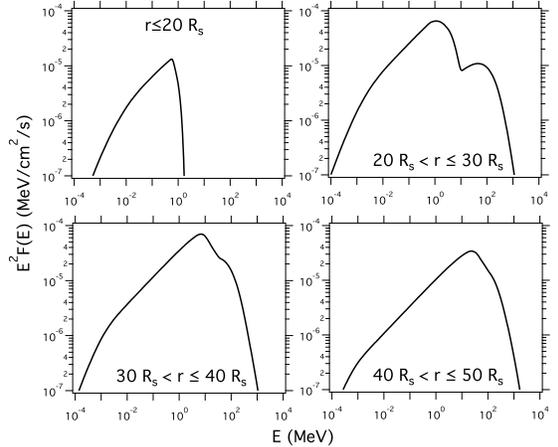}
\caption{The spectra of different regions where the original curvature photons are emitted.}
\label{Spc_section}
\end{figure}

\begin{figure}[h]
\plotone{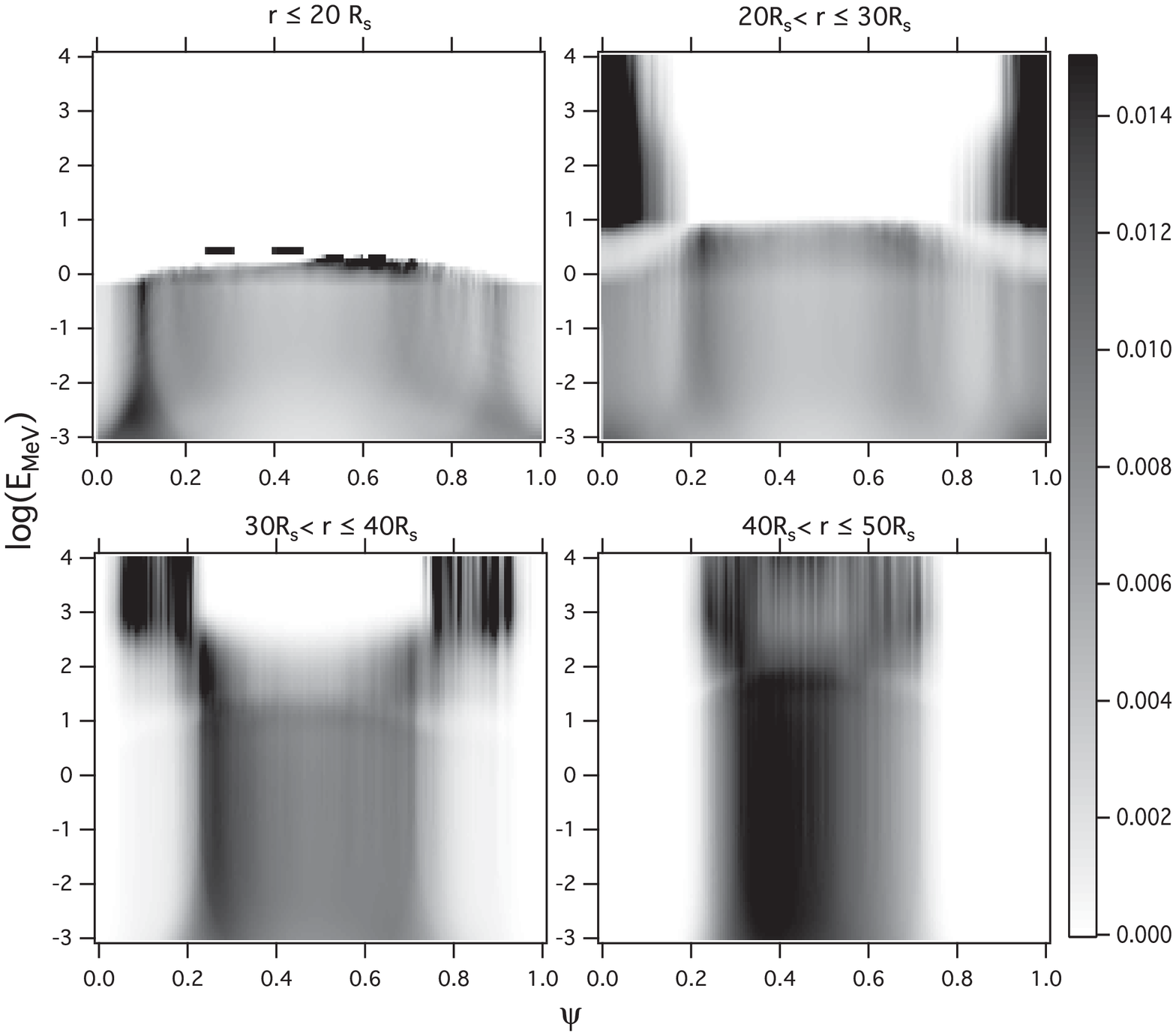}
\caption{The energy dependent light curves of different regions where the original curvature photons are emitted. The darkness represents the percentage of the number of photons of a certain pulse phase interval in the total number of photons of a certain energy range.}
\label{Colmap_section}
\end{figure}

The observations (Kuiper et al. 1999; Abdo et al. 2010; Pilia et al. 2010) indicate that there is another peak around $\psi\sim{}0$ in the light curve when $E>10$MeV. Here the energy dependent light curves (Figure~\ref{LC_section}, Figure~\ref{Colmap_section}), phase averaged spectra (Figure~\ref{Spc_section}) and the skymaps of the synchrotron radiation (Figure~\ref{Skymap_section}) of the regions with different distances from the star under $\alpha=20^{\circ}$ and $\beta=11^{\circ}$ are made to explain this second peak. The reason that the emission region is separated by the distance from the star, is that the dipole magnetic field plays an important role in the resultant shape of the light curve of the synchrotron radiation, by determining the pitch angles of the secondary pairs, which are generated via the magnetic pair creation. The pitch angle correlates with the distance between the place where the pair creation happens and the birth place of the original curvature photons. The farther the curvature photon goes before being converted to a pair, the larger the pitch angle the secondary pair has. According to Equation~(\ref{appro_tauB}), the distance of the curvature photon's penetration is mainly determined by two factors: the magnetic field along the path of the curvature photon, and the energy of the photon. For a fixed $B_{\perp}$ and $E_{\gamma}$, if the $B(\vec{r})$ is higher, then the corresponding $\theta_p(\vec{r})$ is smaller, which means a shorter distance between the place of pair creation and the birth place of the original curvature photon. And under a fixed $B(\vec{r})$, the magnetic pair creation of the photon with a higher energy $E_{\gamma}$ requires a smaller pitch angle $\theta_p(\vec{r})$.

To show how the magnetic field $B(\vec{r})$ changes the shape of the light curve by changing $\theta_p(\vec{r})$, we make the light curves of four energy bands, which are shown in Figure~\ref{LC_section}, whose original curvature photons are generated in five regions: $r\le{}20R_s$, $20R_s<r\le{}30R_s$, $30R_s<r\le{}40R_s$, $40R_s<r\le{}50R_s$, and $r>50R_s$. This figure shows that the X-ray photons' energy originates from particles of $r\le{}50R_s$, of which the region of $30R_s<r\le{}50R_s$ provides the asymmetric wide single peak, and the region of $20R_s<r\le{}30R_s$ provides a background that smears out the little peak of $\psi\sim{}0.65$. For the soft $\gamma$-rays, the two components of the light curves of $E>10$MeV: an asymmetric wide single peak at $\psi\sim{}0.25-0.65$ and a sharp peak at $\psi\sim{}0$, correspond to the regions of $30R_s<r\le{}50R_s$ and $20R_s<r\le{}30R_s$, respectively. The magnetic fields with different strengths generate distinct pulse profiles of the synchrotron radiation of the secondary pairs.

Why is there a peak at $\psi\sim0$? Suppose there are some curvature photons emitted by the particles moving along the field lines towards the stellar surface at the place $\vec{r}$, on their path along $\vec{v}_{cur}$ given by Equation~(\ref{direct_cur}), then the photons with higher energy become pairs earlier, which means the pairs have smaller pitch angles $\theta_p$. In Figure~\ref{two_lines}, the cone of the synchrotron radiation with a small $\theta_p$ is a ``circle" around the polar cap, while the cone of a large $\theta_p$ is a ``line" going across the skymap. When $\beta<\alpha$, such ``circles'' contribute the observed peak at $\psi\sim{}0$ when $E>10$MeV, and the ``lines'' make the wide single peak. As shown in Figure~\ref{Skymap_section}, when $r\le{}30R_s$, there are such ``circles'' around the polar cap, while when $r>30R_s$, there are only ``lines'', which means these synchrotron cones have large pitch angles. This is because the magnetic field is stronger at $r\le{}30R_s$, and, as a result, the photons with higher energy can become pairs under smaller $\theta_p$. For $r>30R_s$, without enough strength of the magnetic field, most of the photons need large pitch angles ($\le{}90^{\circ}$) to satisfy the condition of pair creation, so in the skymap, there is no component around the polar cap.

The attenuation of the synchrotron photons caused by the magnetic field also plays an important role in making the second peak. For an individual synchrotron cone at $\vec{r'}$, some of the emitted photons still have enough energy to become pairs when they approach the stellar surface. Although this process is not determined by the birth place of the curvature photons directly, it can be found in Figure~\ref{Spc_section} that the spectra of different $r$ have different levels of attenuation of the synchrotron radiation. For the case of $r\le{}20R_s$,  few of the synchrotron photons with $E>1$MeV survive from the magnetic pair creation. As $r$ increases, more and more synchrotron photons can survive, when $r>40R_s$, the attenuation is already insignificant. The spectrum of the survival synchrotron radiation of a certain range of $r$ has two components: a spectrum without experiencing the pair creation, and a partially absorbed spectrum. The unabsorbed component cuts off at the maximum energy of the survival photons, which corresponds to the maximum magnetic field they experienced along their paths $\vec{v}_{syn}$. This maximum magnetic field becomes smaller as the birth place of the curvature photon $\vec{r}$ gets further away from the stellar surface. These spectra are consistent with the energy dependent light curves that are shown in Figure~\ref{Colmap_section}. From these light curves, it is found that the synchrotron radiation, whose pulse phase is closer to $\psi\sim{}0$, has a higher chance to survive from the magnetic pair creation. This is because the synchrotron radiation with smaller pitch angle is emitted further from the stellar surface, where the photons have a higher chance to survive.

Why does the peak at $\psi\sim0$ show up when the energy increases? If the particle is made via magnetic pair creation, its typical synchrotron energy $E_{syn}$ and the typical energy of the original curvature photon $E_{cur}$ nearly satisfy $E_{syn}\propto{}E_{cur}$, which can be obtained by combining Equation~(\ref{Esyn_typ}) and (\ref{appro_tauB}). Higher $E_{cur}$ leads to smaller $\theta_p$. Therefore, for the curvature photons emitted from $\vec{r}$, under $\beta<\alpha$, the observed spectrum of the synchrotron radiation, whose pulse phase is closer to $\psi=0$, has a higher cut-off energy.

To summarize, for a fixed birth place of the curvature photons $\vec{r}$, the photons with higher $E_{cur}$ become pairs earlier with smaller $\theta_p$ and further from the star, which raises the synchrotron photons' chances to avoid from becoming pairs. For a curvature photon with $E_{cur}$, if it is emitted further from the star, the $\theta_p$ of the secondary pair is larger, because of the smaller magnetic field. When $\beta<\alpha$, the pulse phase of the synchrotron radiation is closer to 0 when the pitch angle is smaller, and closer to 0.5 when the pitch angle is larger.

Therefore, the second peak of PSR B1509 is the synchrotron radiation emitted by the pairs with smaller pitch angles, which are converted from the ones with higher energy of the curvature photons emitted from the region of $20R_s<r<30R_s$. The viewing angle $\beta$ determines the energy where the second peak shows up. If $\beta<\alpha$, the second peak shows up at higher energy when $\beta$ gets closer to $\alpha$, because the observed synchrotron radiation is emitted by the pairs with smaller pitch angle $\theta_p$, which were the curvature photons of higher energy. The region of $r>30R_s$ contributes the wide single peak of $\psi\sim{}0.25-0.65$. Because each synchrotron cone sweeps the line of sight twice on average, in the light curve, there are two peaks besides $\psi\sim{}0.5$, which overlap each other, making a wide single peak. The rotation of the neutron star make this single peak asymmetric.

\section{Summary}
The outer gap model predicts that most pairs created inside the gap are around the null charge surface and the electric field separates the opposite charges to move in opposite directions. Consequently, from the null charge surface to the light cylinder the outflow radiation is dominant whereas from the null charge surface to the star the inflow radiation is dominant. Since the electric field decreases rapidly from the null charge surface to the star, the incoming radiation flux is of an order of magnitude weaker than that of the outgoing flux. Furthermore most of the incoming high energy curvature photons are converted into pairs by the strong stellar magnetic field of the canonical pulsars. Based on these model features we propose a model to calculate the non-thermal X-rays and soft $\gamma$-rays emitted by secondary pairs produced by converting the incoming curvature photons emitted by the outer gap's primary charged particles. This model is used to explain the observed X-ray and soft $\gamma$-ray spectra and the energy dependent light curves of PSR B1509 from 100eV to GeV. We argue that the line of sight of PSR B1509 is in the direction of the incoming beam instead of the outgoing beam, otherwise a characteristic power law with exponential cut-off spectrum with a cut-off energy around a few GeV should be observed.

The secondary pairs spiral around the magnetic field and emit synchrotron photons, some of these synchrotron photons can also become pairs via magnetic pair creation. In order to avoid seeing the outgoing flux and fit the observed multi-wavelength light curves we need to choose $\alpha=20^{\circ}$ and $\beta=11^{\circ}$. The observed spectrum is the synchrotron radiation emitted from the pairs produced by the magnetic field that converts the major part of the incoming curvature photons. The peak position of the observed spectrum naturally occurs around one MeV because photons above this energy scale can be converted into pairs by the strong stellar magnetic field if they get close enough to the star.

The magnetic pair creation requires a significant pitch angle of the photon. For the curvature photons emitted from a certain place, the magnetic field and their pitch angles increase along their path, the ones with higher energy become pairs first, which have smaller pitch angles and higher cut-off energies of their spectra, while the ones with lower energy need to go further inwards to obtain large enough perpendicular magnetic field, and they become pairs with larger pitch angles and lower cut-off energies. The photons emitted by the pairs with larger pitch angles have a higher chance to become pairs further on.
Therefore the differences between the light curves of different energy bands are due to the different pitch angles of the secondary pairs, and the second peak appearing at $E>10$MeV comes from the region near the star, where the stronger magnetic field allows pair creation to happen with a smaller pitch angle.
 
Although in this paper our model is applied to explain the soft $\gamma$-ray spectrum and the energy dependent light curves from 100 eV to GeV of PSR B1509,  it can be extended to other pulsars as a general framework for studying the non-thermal X-ray and soft $\gamma$-ray emissions.

\section*{Acknowledgments}
We thank Alice K. Harding and W. Hermsen for useful discussion, and thank K. MacKeown for the critical reading. This work is supported by a GRF grant of the Hong Kong Government under 700911P.

\end{document}